\documentclass[twocolumn,showpacs,aps,prl,superscriptaddress,preprintnumbers,amsmath,amssymb,floatfix,linenumbers]{revtex4}

\usepackage{graphicx}    
\usepackage{dcolumn}     
\usepackage{bm}          
\usepackage{color}

\newcommand{\BABARPubYear}    {13}
\newcommand{\BABARPubNumber}  {005}

\newcommand{\SLACPubNumber}   {15448}

\input babarsym
\newcommand{\bi}{\begin{itemize}}
\newcommand{\ei}{\end{itemize}}
\newcommand{\ben}{\begin{enumerate}}
\newcommand{\een}{\end{enumerate}}
\newcommand{\bc}{\begin{center}}
\newcommand{\ec}{\end{center}}
\newcommand{\bt}{\begin{table}}
\newcommand{\et}{\end{table}}
\newcommand{\be}{\begin{equation}}
\newcommand{\eeq}{\end{equation}}
\newcommand{\ba}{\begin{eqnarray}}
\newcommand{\ea}{\end{eqnarray}}

\newcommand{\la}{\ifmmode {\leftarrow} \else {$\leftarrow$}\fi}
\newcommand{\Ra}{\ifmmode {\Rightarrow} \else {$\Rightarrow$}\fi}
\newcommand{\La}{\ifmmode {\Leftarrow} \else {$\Leftarrow$}\fi}
\newcommand{\Lra}{\ifmmode {\Longrightarrow} \else {$\Longrightarrow$}\fi}
\newcommand{\Lla}{\ifmmode {\Longleftarrow} \else {$\Longleftarrow$\fi}}
\newcommand{\Llra}{\ifmmode {\Longleftrightarrow} \else {$\Longleftrightarrow$\fi}}
\newcommand{\Lk}{\ifmmode {{\cal L}} \else {${\cal L}$}\fi}
\newcommand{\Wt}{\ifmmode {{\cal W}} \else {${\cal W}$}\fi}
\newcommand{\Br}{\ifmmode {{\cal B}} \else {${\cal B}$}\fi}
\newcommand{\N}{\ifmmode {{\cal N}} \else {${\cal N}$}\fi}
\newcommand{\G}{\ifmmode {{\cal G}} \else {${\cal G}$}\fi}
\newcommand{\E}{\ifmmode {{\cal E}} \else {${\cal E}$}\fi}
\newcommand{\Pfr}{\ifmmode {{\cal F}} \else {${\cal F}$}\fi}
\newcommand{\Aone}{\ifmmode {{\cal A}_1} \else {${\cal A}_1$}\fi}
\newcommand{\rha}{\ifmmode{\mbox{\rho^2_{{\cal A}_1}}} \else {\mbox{$\rho^2_{{\cal A}_1}$}}\fi}
\newcommand{\rhf}{\ifmmode{\rho^2_{\cal F}}\else{\mbox{$\rho^2_{\cal F}$}}\fi}
\newcommand{\om}{\ifmmode {w} \else {$w$}\fi}
\newcommand{\dom}{\ifmmode {\Delta w} \else {$\Delta w$}\fi}

\newcommand{\tBz}{\ifmmode {\tau_{\Bz}} \else {$\tau_{\Bz}$}\fi}
\newcommand{\tBp}{\ifmmode {\tau_{\Bu}} \else {$\tau_{\Bu}$}\fi}

\newcommand{\psoft}{\ifmmode {\pi_{s}^-} \else {$\pi_{s}^-$}\fi}
\newcommand{\plab}{\ifmmode{p} \else {$p$} \fi}
\newcommand{\ctdl}{\ifmmode{ \cos(\theta_{\Dstar\ell}) } \else {$\cos(\theta_{\Dstar\ell})$} \fi}
\newcommand{\ks}{\ifmmode{k^*} \else {$k^*$} \fi}
\newcommand{\mnutag}{\ifmmode{m^2_{\nu ,tag}} \else {$m^2_{\nu ,tag}$}\fi} 
\newcommand{\mnusig}{\ifmmode{m^2_{\nu ,sig}} \else {$m^2_{\nu ,sig}$}\fi} 
\newcommand{\DTau}{\ifmmode {\Delta \tau} \else {$\Delta \tau$}\fi}
\newcommand{\ggcc}{\ifmmode {GeV^2/c^4} \else {$GeV^2/c^4$}\fi}
\def\BpBm {\ensuremath{B^+ {\kern -0.16em \Bub}}}

\def\dT{\ensuremath{\Delta t}}
\def\DT{\ensuremath{\Delta t}}

\def\poverq2{\ensuremath{\bigg\vert\frac{p}{q}\bigg\vert^2}\xspace}
\def\qoverp2{\ensuremath{\bigg\vert\frac{q}{p}\bigg\vert^2}\xspace}

\def\BzBz     {\ensuremath{\mbox{\Bz {\kern -0.1em \Bz}}}\xspace}
\def\BzBzb    {\ensuremath{\mbox{\Bz {\kern -0.1em \Bzb}}}\xspace}
\def\BzbBzb   {\ensuremath{\mbox{\Bzb {\kern -0.1em \Bzb}}}\xspace}

\newcommand{\Brec}{\mbox{$B_{R}$}}
\newcommand{\Btag}{\mbox{$B_{T}$}}

\def\Kp{\ensuremath {K^{+}\xspace}}
\def\Km{\ensuremath {K^{-}\xspace}}
\def\Kt{\ensuremath {K_{tag}\xspace}}
\def\Kr{\ensuremath {K_{rec}\xspace}}

\def\At{\ensuremath{\mathcal{A}_{K}}\xspace}
\def\Ar{\ensuremath{\mathcal{A}_{r\ell}}\xspace}

\def\All{\ensuremath{\mathcal{A}_{CP}}\xspace}

\def\Kt{\ensuremath{K_{T}}\xspace}
\def\Kr{\ensuremath{K_{R}}\xspace}

\def\dCP{\ensuremath{\Delta_{CP}}}

\def\qoverp{\ensuremath{\frac{q}{p}}}

\newcommand{\dmd}{\mbox{$\Delta m_d$}}

\long\def\inst#1{\par\nobreak\kern 4pt\nobreak
    {\it #1}\par\vskip 10pt plus 3pt minus 3pt}
\RequirePackage{xspace}

\usepackage{relsize}
\def\babar{\mbox{\slshape B\kern-0.1em{\smaller A}\kern-0.1em
    B\kern-0.1em{\smaller A\kern-0.2em R}}}

\def\en         {\ensuremath{e^-}\xspace}   
\def\ep         {\ensuremath{e^+}\xspace}

\def\mup        {\ensuremath{\mu^+}\xspace}
\def\mun        {\ensuremath{\mu^-}\xspace} 
\def\mumu       {\ensuremath{\mu^+\mu^-}\xspace}

\def\taup       {\ensuremath{\tau^+}\xspace}
\def\taum       {\ensuremath{\tau^-}\xspace}

\def\ellm       {\ensuremath{\ell^-}\xspace}
\def\ellp       {\ensuremath{\ell^+}\xspace}

\def\ellell     {\ensuremath{\ell^+ \ell^-}\xspace}
\def\nub        {\ensuremath{\overline{\nu}}\xspace}
\def\nunub      {\ensuremath{\nu{\overline{\nu}}}\xspace}
\def\nub        {\ensuremath{\overline{\nu}}\xspace}
\def\nunub      {\ensuremath{\nu{\overline{\nu}}}\xspace}
\def\nue        {\ensuremath{\nu_e}\xspace}
\def\nueb       {\ensuremath{\nub_e}\xspace}

\def\num        {\ensuremath{\nu_\mu}\xspace}
\def\numb       {\ensuremath{\nub_\mu}\xspace}

\def\nut        {\ensuremath{\nu_\tau}\xspace}
\def\nutb       {\ensuremath{\nub_\tau}\xspace}

\def\nul        {\ensuremath{\nu_\ell}\xspace}
\def\nulb       {\ensuremath{\nub_\ell}\xspace}

\def\g     {\ensuremath{\gamma}\xspace}
\def\gaga  {\ensuremath{\gamma\gamma}\xspace}  
\def\ggstar{\ensuremath{\gamma\gamma^*}\xspace}

\def\piz   {\ensuremath{\pi^0}\xspace}

\def\ppz   {\ensuremath{\pi^0\pi^0}\xspace}
\def\pip   {\ensuremath{\pi^+}\xspace}
\def\pim   {\ensuremath{\pi^-}\xspace}
\def\pipi  {\ensuremath{\pi^+\pi^-}\xspace}

\def\pimp  {\ensuremath{\pi^\mp}\xspace}
\def\kaon  {\ensuremath{K}\xspace}
\def\Kbar  {\kern 0.2em\overline{\kern -0.2em K}{}\xspace}

\def\Kz    {\ensuremath{K^0}\xspace}
\def\Kzb   {\ensuremath{\Kbar^0}\xspace}
\def\KzKzb {\ensuremath{\Kz \kern -0.16em \Kzb}\xspace}
\def\Kp    {\ensuremath{K^+}\xspace}
\def\Km    {\ensuremath{K^-}\xspace}
\def\Kpm   {\ensuremath{K^\pm}\xspace}

\def\KpKm  {\ensuremath{\Kp \kern -0.16em \Km}\xspace}
\def\KS    {\ensuremath{K^0_{\scriptscriptstyle S}}\xspace} 
\def\KL    {\ensuremath{K^0_{\scriptscriptstyle L}}\xspace} 
\def\Kstarz  {\ensuremath{K^{*0}}\xspace}

\def\Kstar   {\ensuremath{K^*}\xspace}

\def\Kstarp  {\ensuremath{K^{*+}}\xspace}

\def\Dbar    {\kern 0.2em\overline{\kern -0.2em D}{}\xspace}

\def\Dz      {\ensuremath{D^0}\xspace}
\def\Dzb     {\ensuremath{\Dbar^0}\xspace}
\def\DzDzb   {\ensuremath{\Dz {\kern -0.16em \Dzb}}\xspace}
\def\Dp      {\ensuremath{D^+}\xspace}
\def\Dm      {\ensuremath{D^-}\xspace}

\def\DpDm    {\ensuremath{\Dp {\kern -0.16em \Dm}}\xspace}
\def\Dstar   {\ensuremath{D^*}\xspace}

\def\Dstarz  {\ensuremath{D^{*0}}\xspace}

\def\Dstarp  {\ensuremath{D^{*+}}\xspace}
\def\Dstarm  {\ensuremath{D^{*-}}\xspace}

\def\B       {\ensuremath{B}\xspace}
\def\Bbar    {\kern 0.18em\overline{\kern -0.18em B}{}\xspace}
\def\Bb      {\ensuremath{\Bbar}\xspace}
\def\BB      {\ensuremath{\B {\kern -0.16em \Bb}}\xspace}
\def\BzBzb   {\ensuremath{\Bz {\kern -0.16em \Bzb}}\xspace}

\def\Bu      {\ensuremath{B^+}\xspace}
\def\Bub     {\ensuremath{B^-}\xspace}
\def\Bp      {\ensuremath{\Bu}\xspace}

\def\Bpm     {\ensuremath{B^\pm}\xspace}

\def\BpBm    {\ensuremath{\Bu {\kern -0.16em \Bub}}\xspace}
\def\Bs      {\ensuremath{B_s}\xspace}

\def\BorBbar    {\kern 0.18em\optbar{\kern -0.18em B}{}\xspace}
\def\DorDbar    {\kern 0.18em\optbar{\kern -0.18em D}{}\xspace}
\def\KorKbar    {\kern 0.18em\optbar{\kern -0.18em K}{}\xspace}

\def\jpsi     {\ensuremath{{J\mskip -3mu/\mskip -2mu\psi\mskip 2mu}}\xspace}

\mathchardef\Upsilon="7107
\def\Y#1S{\ensuremath{\Upsilon{(#1S)}}\xspace}

\def\FourS {\Y4S}

\mathchardef\Deltares="7101
\mathchardef\Xi="7104
\mathchardef\Lambda="7103
\mathchardef\Sigma="7106
\mathchardef\Omega="710A

\def\Deltabar{\kern 0.25em\overline{\kern -0.25em \Deltares}{}\xspace}
\def\Lbar{\kern 0.2em\overline{\kern -0.2em\Lambda\kern 0.05em}\kern-0.05em{}\xspace}
\def\Sigbar{\kern 0.2em\overline{\kern -0.2em \Sigma}{}\xspace}
\def\Xibar{\kern 0.2em\overline{\kern -0.2em \Xi}{}\xspace}
\def\Obar{\kern 0.2em\overline{\kern -0.2em \Omega}{}\xspace}
\def\Nbar{\kern 0.2em\overline{\kern -0.2em N}{}\xspace}
\def\Xb{\kern 0.2em\overline{\kern -0.2em X}{}\xspace}
\def\X {\ensuremath{X}\xspace}

\def\bpsikl     {\ensuremath{\Bz \to \jpsi \KL}\xspace}

\def\Bzbtox     {\ensuremath{\Bzb \to \X}\xspace}

\def\Bztodstlnu  {\ensuremath{\Bz \to \dsm X \ellp \nu}\xspace}

\def\invfb   {\ensuremath{\mbox{\,fb}^{-1}}\xspace}

\def\mus  {\ensuremath{\rm \,\mus}\xspace}
\def\ns   {\ensuremath{\rm \,ns}\xspace}
\def\ps   {\ensuremath{\rm \,ps}\xspace}

\def\mus        {\ensuremath{\,\mu{\rm s}}\xspace}    
\def\ns         {\ensuremath{{\rm \,ns}}\xspace}      
\def\ps         {\ensuremath{{\rm \,ps}}\xspace}  

\def\degrees{\ensuremath{^{\circ}}\xspace}

\def\psoft {\ensuremath{{\pi_s^-}}\xspace}    

\def\order{{\ensuremath{\cal O}}\xspace}

\def\to                 {\ensuremath{\rightarrow}\xspace}
\def\pep2{PEP-II}
\def\BF{$B$ Factory}

\def\gsim{{~\raise.15em\hbox{$>$}\kern-.85em
          \lower.35em\hbox{$\sim$}~}\xspace}
\def\lsim{{~\raise.15em\hbox{$<$}\kern-.85em
          \lower.35em\hbox{$\sim$}~}\xspace}

\def\CP                {\ensuremath{C\!P}\xspace}

\def\epstag  {\ensuremath{\varepsilon_{\rm tag}}\xspace}

\def\deltaz{\ensuremath{{\rm \Delta}z}\xspace}
\def\deltat{\ensuremath{{\rm \Delta}t}\xspace}
\def\deltamd{\ensuremath{{\rm \Delta}m_d}\xspace}

\def\jetset74   {\mbox{\tt Jetset \hspace{-0.5em}7.\hspace{-0.2em}4}\xspace}

\def\qop {\ensuremath{{|q/p|}}\xspace}
\def\ACP {\ensuremath{{\cal A}_{CP}}\xspace}
\def\All {\ensuremath{{\cal A}_{\ell\ell}}\xspace}

\def\Brec {\ensuremath{B_{R}}\xspace}
\def\Btag {\ensuremath{B_{T}}\xspace}

\usepackage{relsize}

\def\Mnu{\ensuremath{{\cal M}_\nu^2}}

\begin{document}

\begin{flushleft}
{\babar-PUB-\BABARPubYear/\BABARPubNumber \\
SLAC-PUB-\SLACPubNumber     \\
}
\end{flushleft}

\title{Search for {\boldmath \CP} Violation in {\boldmath \BzBzb} Mixing
using Partial Reconstruction of
{\boldmath{$\Bz \rightarrow \dsm X \ell^{+} \nu_{\ell}$}}
and a Kaon Tag}

%
\author{J.~P.~Lees}
\author{V.~Poireau}
\author{V.~Tisserand}
\affiliation{Laboratoire d'Annecy-le-Vieux de Physique des Particules (LAPP), Universit\'e de Savoie, CNRS/IN2P3,  F-74941 Annecy-Le-Vieux, France}
\author{E.~Grauges}
\affiliation{Universitat de Barcelona, Facultat de Fisica, Departament ECM, E-08028 Barcelona, Spain }
\author{A.~Palano$^{ab}$ }
\affiliation{INFN Sezione di Bari$^{a}$; Dipartimento di Fisica, Universit\`a di Bari$^{b}$, I-70126 Bari, Italy }
\author{G.~Eigen}
\author{B.~Stugu}
\affiliation{University of Bergen, Institute of Physics, N-5007 Bergen, Norway }
\author{D.~N.~Brown}
\author{L.~T.~Kerth}
\author{Yu.~G.~Kolomensky}
\author{M.~J.~Lee}
\author{G.~Lynch}
\affiliation{Lawrence Berkeley National Laboratory and University of California, Berkeley, California 94720, USA }
% ...
\collaboration{The \babar\ Collaboration}
\noaffiliation

\begin{abstract}
We present results of a search for \CP violation in \BzBzb mixing
with the \babar\ detector. We select a sample of \Bztodstlnu  decays
with a partial reconstruction method and use kaon tagging to assess the
flavor of the other \B meson in the event. 
We determine the \CP violating asymmetry
 $\ACP \equiv \frac{N(\Bz\Bz) - N(\Bzb\Bzb)}{N(\Bz\Bz)+ N(\Bzb\Bzb)}
=(0.06\pm 0.17^{+0.38}_{-0.32})\%$, 
corresponding to
$\dCP =1 - |q/p| =(0.29\pm0.84^{+1.88 }_{-1.61})~\times~10^{-3}$.

\end{abstract}


\maketitle

Experiments at \B factories have observed \CP violation 
in direct \Bz decays \cite{directCP} and 
in the interference between \Bz mixing and decay \cite{sin2b}. 
\CP violation in mixing has so far eluded observation.

The weak-Hamiltonian
eigenstates are related to the flavor eigenstates of the strong
interaction Hamiltonian 
by $| B_{L,H}\rangle = p| \Bz\rangle \pm q |\Bzb\rangle$. 
The value of the ratio \qop can be determined from the asymmetry
between the two oscillation probabilities ${\cal P} = P(\Bz \rightarrow
\Bzb)$ and 
$\bar{\cal P} = P(\Bzb \rightarrow \Bz)$ through 
$\ACP = (\bar{{\cal P}} -{\cal P})/(\bar{{\cal P}} +{\cal P}) 
= \frac{1-\qop^4}{1+\qop^4} \approx 2\dCP $, 
where $\dCP = 1 -|q/p|$ and the
Standard Model (SM) prediction is
$\ACP= -(4.0 \pm 0.6) \times 10^{-4}$~\cite{Nierste}.
Any observation with the present
experimental sensitivity ($\order(10^{-3})$) would therefore reveal
physics beyond the SM.

Experiments 
measure \ACP from the dilepton asymmetry, 
$\All = \frac{N(\ellp\ellp)-N(\ellm\ellm)}
{N(\ellp\ellp)+N(\ellm\ellm)} $, where an~\ellp
(\ellm) 
tags a \Bz (\Bzb) meson, and $\ell$ refers either to an electron or a muon~\cite{epem_dilep}.
These measurements benefit from the large number of produced dilepton events. 
However, they rely on
the use of control samples to subtract the charge-asymmetric
background originating from
hadrons wrongly identified as leptons
or leptons from light hadron
decays, and to compute the charge-dependent lepton identification
asymmetry that may produce a false signal. The systematic
uncertanties associated with the corrections for these effects 
constitute a severe limitation to the precision 
of the measurements.

Using a sample of dimuon events, the $D\emptyset$ Collaboration 
 measured a value of \ACP for a mixture of \Bs
  and \Bz decays that deviates from the SM by 3.9 standard 
  deviations  ~\cite{D0_mumu}. 
Measurements of \ACP for \Bs mesons
performed by the $D\emptyset$ Collaboration
with $\Bs \rightarrow D_{s} \mu X$ decays are consistent with 
the SM~\cite{D0_Bs}.



We present a measurement of \ensuremath{{\cal A}_{CP}(B^0)} with a new analysis technique. 
We reconstruct
a sample of \Bz mesons (hereafter called \Brec; charge conjugate states are
implied unless otherwise stated) from the
semileptonic transition \Bztodstlnu ,
with a partial reconstruction of the $\dsm \rightarrow \pi^- \Dbar^0$
decay (see Ref.~\cite{DKpi} and references therein).  
The observed asymmetry between the number 
of events with an $\ell^+$ compared to those with an $\ell^-$ is then:
\begin{eqnarray}
\label{eq:arec}   A_\ell \approx \Ar + \ACP  \chi_d,
\end{eqnarray}
where $\chi_d=0.1862\pm0.0023$~\cite{PDG} is the integrated mixing 
probability for \Bz\
mesons and \Ar\ is the detector-induced 
charge asymmetry in the \Brec reconstruction.

We identify (``tag'') the flavor of the other \Bz meson (labeled
\Btag) using events with a charged kaon (\Kt).
An event with a \Kp~(\Km) usually arises from a state that decays as a
\Bz~(\Bzb) meson. When mixing takes places, the $\ell$ and the \Kt\ then
have the same electric charge. 
The observed asymmetry in the rate of mixed events is:
\begin{eqnarray}
\label{eq:amix}   A_{T} = \frac{N(\ell^+\Kt^+) - N(\ell^-\Kt^-)} {N(\ell^+\Kt^+) + N(\ell^-\Kt^-)} \approx \Ar + \At + \ACP ,
\end{eqnarray}
where \At\ is the detector charge asymmetry in kaon reconstruction.
A kaon with the same charge as the $\ell$
might also arise from the Cabibbo-Favored (CF) decays  of the \Dz\
meson produced with the lepton from the partially reconstructed side 
(\Kr).
The asymmetry observed for these events is:
\begin{eqnarray}
\label{eq:asame} A_{R} =\frac{N(\ell^+\Kr^+) - N(\ell^-\Kr^-)} {N(\ell^+\Kr^+) + N(\ell^-\Kr^-)} \approx \Ar+\At + \ACP \chi_d.
\end{eqnarray}

Eqs.~\ref{eq:arec}, \ref{eq:amix}, and \ref{eq:asame} can be
used to extract \ACP\ and the detector induced asymmetries (\Ar and \At).

A detailed description of the \babar\ detector is provided 
elsewhere~\cite{babar_nim}. 
We use a sample with an integrated luminosity of 425.7 \invfb
~\cite{babar_lumi} collected 
on the peak of
the \FourS resonance. A 45 \invfb sample collected 40\mev below the
resonance (``off-peak'')  is used for background studies. 
We also use
a simulated sample of $\BB$ events~\cite{evtgen} with an integrated luminosity equivalent to approximately 
three times the data.

We preselect a sample of hadronic events requiring the number of 
charged particles to
be at least four.
We reduce non-\BB (continuum) background 
by requiring the ratio 
of the second to the zeroth order Fox-Wolfram moments~\cite{wolfram} to be less
than 0.6.

We select the \Brec\ sample by
searching for 
combinations of a charged lepton
(in the momentum range $1.4 < p_\ell < 2.3$ \gevc
) and a low
 momentum pion \psoft ($60 < p_\psoft < 190$ \mevc), which is
 taken to arise from $\dsm \to \Dbar^0 \psoft$ decay. 
Here and elsewhere momenta are calculated in the-center-of-mass frame.
The \ellp and the \psoft must have opposite electric charge. Their
tracks must be consistent with originating from a common vertex, which
is constrained to the 
beam collision point in the plane transverse to the beam axis. Finally, we combine $p_{\ell}$, $p_{\psoft}$, and the
probability of the vertex fit in a likelihood ratio variable ($\eta$) optimized to reject combinatorial \BB\ events. 
If more than one candidate is found in the event, we choose
the one with the largest value of $\eta$.

We determine the square of the unobserved neutrino mass as:
\begin{eqnarray}
\nonumber
\Mnu = (E_{\mbox{\rm beam}}-E_{{D^*}} - 
E_{\ell})^2-({\bf{p}}_{{D^*}} + {\bf{p}}_{\ell})^2 ,
\label{eqn:mms}
\end{eqnarray}
where we neglect the momentum of the \Bz\ 
(p$_B$ \mbox{$\approx$ 340 \mevc}) and identify the \Bz\ energy with the beam 
energy $E_{\mbox{\rm beam}}$ in the $e^+e^-$ center-of-mass frame; 
$E_{\ell}$ and  ${\bf{p}}_{\ell}$ are the energy and momentum of the 
lepton and ${\bf{p}}_{{D^*}}$ is the estimated momentum of the $D^*$.
As a consequence of the limited phase space available in the $D^{*+}$
decay, the soft pion is emitted nearly at rest in the $D^{*+}$ rest frame.
The $D^{*+}$ four-momentum can therefore be computed by approximating 
its direction as that of the soft pion, and parametrizing its momentum as 
a linear function of the soft-pion momentum.
All \Bz\ semileptonic decays with \Mnu\ near zero are considered
to be signal events, including  
$\Bz \rightarrow \dsm X^0 \ell^+ \nu_{\ell}$ 
(primary), 
$\Bz \rightarrow \dsm X^0 \tau^+ \nu_{\tau}, ~\tau^+ \rightarrow \ellp
\nu_{\ell} \bar{\nu}_\tau$ 
(cascade), and $\Bz \rightarrow \dsm h^+$ (misidentified), where the hadron ($h = \pi,K)$
is erroneously identified as a lepton (in most cases, a muon). 
\Bz decays to flavor-insensitive 
\CP eigenstates, $\Bz \rightarrow D^{*\pm} D X, D \rightarrow
\ell^{\mp} X$, and 
$\Bu \rightarrow \dsm X^+ \ell^+ \nu_{\ell}$ 
decays accumulate around zero as the signal events (``peaking background'').
The uncorrelated background consists of continuum and combinatorial \BB\ events.
The latter category includes events where a genuine \dsm\ is combined with an $\ell^+$ from
the other $B$ meson.

We identify charged kaons in the momentum
range $ 0.2 <p_K < 4 $ \gevc with an average
efficiency of about 85\% and a $\sim 3\%$ pion misidentification rate. 
We determine the $K$ production point from the intersection of the 
$K$ track and the beam spot, 
and then
determine the distance \deltaz between the \ellp\psoft and $K$ vertices
coordinates along the beam axis.
Finally, we define
the proper time difference \deltat between the \Brec and the \Btag in the so
called ``Lorentz boost approximation''~\cite{BAPX}, $\deltat = \frac {\deltaz} {\beta \gamma}$,
where the product $\beta\gamma = 0.56$ is the average Lorentz boost of
the \FourS in the laboratory frame.
Since the \B mesons 
are not at rest in the \FourS rest frame,
and in addition the $K$ is usually produced in the cascade process 
$\Btag \rightarrow DX, D \rightarrow K Y$, \deltat is in fact only an
approximation of the actual proper time difference between the \Brec
and the \Btag. 
We reject events if the uncertainty $\sigma(\deltat)$ exceeds 3 ps. 
This selection reduces to a negligible
level the contamination from protons produced in the scattering of
primary particles with the beam pipe or the detector material and
wrongly identified as kaons, which
would otherwise constitute a large charge-asymmetric source of background.

We define an event as ``mixed'' if the $K$ and the $\ell$ have the same electric
charge and as ``unmixed'' otherwise. In about $20\%$ of the cases, the
$K$ has the wrong charge correlation with respect to the \Btag, and the event is
wrongly defined (mistags). 

About 95\% of the \Kr candidates have the same electric charge as the $\ell$;  
they constitute 75\% of the
mixed event sample.
Due to the small
lifetime of the \Dz meson, the separation in space between the
\Kr and the $\ell \pi_{s}$ production points is much smaller 
than for \Kt . Therefore, we use \deltat as a first
discriminant variable.
Kaons  in the \Kr\ sample are usually emitted in the hemisphere opposite to the
$\ell$, while genuine \Kt are produced randomly, so we
use in addition the cosine of the angle $\theta_{\ell K}$ between the $\ell$ and the $K$. 

In about 20\% of the cases,
the events contain more than one $K$; most often we find both 
a \Kt and a \Kr candidate. As these two carry different
information, we accept multiple-candidate events.
Using ensembles of simulated samples of events, we find that this choice does 
not affect the statistical uncertainty. 

\begin{figure}[!htb]
\begin{center}
\includegraphics[width=9cm]{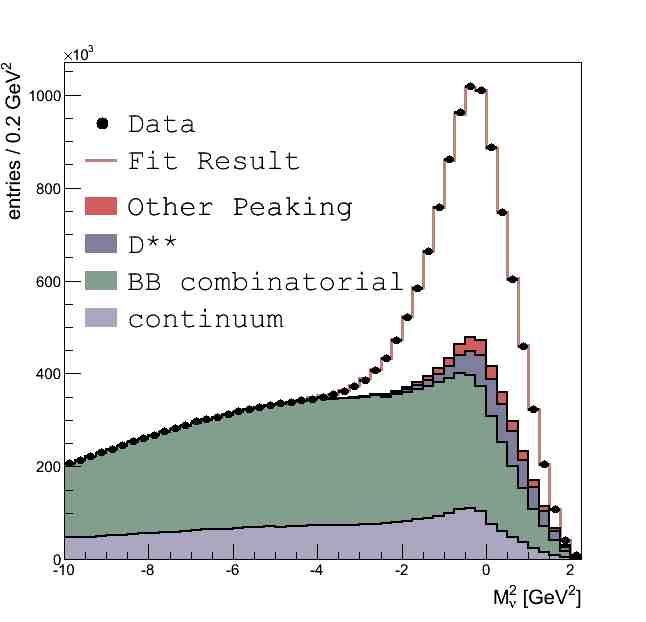}
\caption{(color online). \Mnu\ distribution for selected events. The data are represented 
by the points with error bars. The fitted contributions from 
$\Bzb \rightarrow D^{*+} \ell^{-} \bar\nu_{\ell}$,  
other peaking background, $D^{**}$ events,
  \BB\ combinatorial background, and rescaled off-peak events are overlaid.}
\label{fig:incl_yield}
\end{center}
\end{figure}

The \Mnu\ distribution of all signal candidates in shown in  Fig.~\ref{fig:incl_yield}.
We determine
the signal fraction by fitting the \Mnu\ distribution 
in the interval $[-10, 2.5]~$GeV$^2$/c$^4$ with
the sum of continuum, \BB\ combinatorial, and \BB\ peaking events. 
 We split peaking \BB\ into direct
($\Bz \rightarrow \dsm \ell^+\nu$), 
``\dstrstr'' 
($B \rightarrow \dsm X^0 \ell^+ \nu_{\ell}$), 
cascade, hadrons wrongly identified as leptons, and \CP eigenstates. 
In the fit, we 
float the fraction of direct, \dstrstr, and \BB combinatorial background,
while we fix the continuum contribution to the expectation from 
off-peak events, rescaled by the
on-peak to off-peak luminosity ratio, and
the rest (less than 2\% of the total) to the level predicted by the Monte Carlo
simulation. Based on the assumption of isospin conservation, 
we attribute 66\% of the
\dstrstr events to \Bu decays and the rest to \Bz decays. We use the result
of the fit to compute the fractions of continuum,
combinatorial, and peaking \Bu background, \CP eigenstates, 
and \Bz signal in the sample, 
as a function of \Mnu. We find a total of  $(5.945\pm 0.007) \times 10^6$ 
peaking events (see  Fig.~\ref{fig:incl_yield}).

We then repeat the fit after dividing events in the four lepton categories
($e^\pm , \mu^\pm$) and eight tagged samples 
($e^\pm K^\pm$, $\mu^\pm K^\pm$).

We measure \ACP with a binned four-dimensional fit to \deltat (100
bins), 
$\sigma(\deltat) (20)$, $\cos \theta_{\ell k} (4) $, and $p_{K} (5)$.
Following Ref.~\cite{DCS} and neglecting resolution effects,
the \deltat distributions for signal events with a \Kt are 
represented by the following expressions:
\begin{widetext}
{\small
\nonumber
\begin{eqnarray}
\label{eq:pdf}
\nonumber
\mathcal{F}_{\Bzb \Bz}(\dT) &=&
\frac{\Gamma_0 e^{-\Gamma_0|\dT|}}{2(1+r'^2)}
\bigg[
\left(1+\bigg\vert \frac{q}{p}\bigg\vert^{2} r'^2 \right)\cosh(\Delta\Gamma \dT/2) +
 \left(1-\bigg\vert \frac{q}{p}\bigg\vert^{2} r'^2 \right)\cos(\deltamd \dT) - 
 \bigg\vert{\frac{q}{p}}\bigg\vert  (b+c) \sin(\deltamd \dT)\bigg],  \\
\nonumber
\mathcal{F}_{\Bz \Bzb}(\dT) &=&
\frac{\Gamma_0 e^{-\Gamma_0|\dT|}}{2(1+r'^2)}
\bigg[
\left(1+\bigg\vert \frac{p}{q}\bigg\vert^{2} r'^2 \right)\cosh(\Delta\Gamma \dT/2) + 
 \left(1-\bigg\vert \frac{p}{q}\bigg\vert^{2} r'^2 \right)\cos(\deltamd \dT) + 
 \bigg\vert{\frac{p}{q}}\bigg\vert  (b-c) \sin(\deltamd \dT)\bigg], \\
\nonumber
\mathcal{F}_{\Bzb\Bzb}(\dT) &=& 
\frac{\Gamma_0 e^{-\Gamma_0|\dT|}}{2(1+r'^2)}
\bigg[
\left(1+\bigg\vert \frac{p}{q}\bigg\vert^{2} r'^2 \right)\cosh(\Delta\Gamma \dT/2) - 
 \left(1-\bigg\vert \frac{p}{q}\bigg\vert^{2} r'^2 \right)\cos(\deltamd \dT) - 
 \bigg\vert{\frac{p}{q}}\bigg\vert  (b-c) \sin(\deltamd \dT)\bigg]  
\bigg\vert \frac{q}{p} \bigg \vert^{2}, \\
\nonumber
\mathcal{F}_{\Bz \Bz}(\dT) &=& 
\frac{\Gamma_0 e^{-\Gamma_0|\dT|}}{2(1+r'^2)}
\bigg[
\left(1+\bigg\vert \frac{q}{p}\bigg\vert^{2} r'^2 \right)\cosh(\Delta\Gamma \dT/2) - 
 \left(1-\bigg\vert \frac{q}{p}\bigg\vert^{2} r'^2 \right)\cos(\deltamd \dT) + 
 \bigg\vert{\frac{q}{p}}\bigg\vert  (b+c) \sin(\deltamd \dT)\bigg]
\bigg\vert
\frac{p}{q} \bigg \vert^{2}, 
\nonumber
\end{eqnarray}
}
\end{widetext}
where the first index of ${\mathcal F}$ refers to the flavor of the  \Brec\ 
and the second to the
\Btag , 
$\Gamma_0 = \tau_{\Bz}^{-1}$ is the average width of the two \Bz\
mass eigenstates, 
$\deltamd$ and $\Delta \Gamma$ are respectively their mass and width difference,
the parameter $r'$ results from 
the interference of CF and Doubly Cabibbo
Suppressed (DCS) decays on the \Btag\ side \cite{DCS}
and has a very small value (${\cal O}$(1\%)), and $b$ and $c$ 
are two parameters
expressing the \CP violation arising from that interference.
In the SM, \mbox{$b = 2r' \sin(2\beta+\gamma) \cos\delta'$} and 
 \mbox{$c = -2r' \cos(2\beta+\gamma) \sin\delta'$}, where $\beta$ and
 $\gamma$ are angles of the Unitary Triangle and $\delta'$
 is a strong phase.
The quantities \dmd, \tBz , 
$b$, $c$, and $\sin(2\beta+\gamma)$
are left free in the fit to reduce the systematic uncertainty. 
The value of $\Delta \Gamma$ is fixed to zero.
Neglecting the tiny contribution from DCS decays,
 the main contribution to the asymmetry is time independent and due to
 the normalization factors
 of the two mixed terms.

The \deltat distribution for the decays of the \Bp\ mesons is parametrized by an exponential function, 
 \mbox{${\cal F}_{\Bp} =\Gamma_+ e^{-|\Gamma_+\dT|}$}, where  the \Bp\ decay
width is computed as the inverse of the lifetime 
$\Gamma_+^{-1} = \tau_{B^+} = (1.641\pm 0.008)$ ps.

When the \Kt\ comes from the decay of the \Bz\ meson to a
\CP eigenstate (as, for example $\Bz \rightarrow D^{(*)}\Dbar^{(*)}$~\cite{PDG}),
a different expression applies:
{\small
$$\mathcal{F}_{CPe}(\dT) = \frac{\Gamma_0}{4}
e^{-\Gamma_0|\dT|} [ 1\pm S \sin(\deltamd \dT) \pm C \cos(\deltamd \dT )],$$
}
where the plus sign is used if the \Brec\ decays as a \Bz and the minus sign
otherwise. 
The fraction of these events (about 1\%) and the parameters $S$ and $C$ are
fixed in the fits and are taken from simulation.

We obtain the \DT\ distributions for \Kt in \BB\ events, ${\cal G}_i(\DT)$, 
by convolving the theoretical ones with 
a resolution function, which consists of the superposition of several Gaussian
functions, convolved with exponentials to take into account the finite lifetime
of charmed mesons in the cascade decay $b\rightarrow c\rightarrow K$.
Different sets of
parameters are used for peaking and for combinatorial background events.

To describe the \DT\ distributions for \Kr events, ${\cal G}_{\Kr}(\DT)$,   
we select a
subsample of data containing fewer than 5\% \Kt decays, and use 
background-subtracted histograms in our likelihood functions. 
As an alternative, we apply the same selection to the simulation and 
correct the \DT\ distribution predicted by the Monte Carlo by the
ratio of the histograms extracted from data and simulated events.
The $\cos\theta_{\ell K}$ shapes are obtained 
from the histograms of the
simulated distributions for \BB\ events.
The \DT\ distribution of continuum events is represented by a decaying 
exponential convolved with Gaussians parametrized by fitting simultaneously the
off-peak data.

The rate of events in each bin ($j$) and for each tagged sample are then
expressed as the sum of the predicted contributions from peaking
events, \BB\ combinatorial, and continuum background.
Accounting for mistags and \Kr events, the peaking \Bz\
contributions 
to the same-sign samples are:
{\small
\begin{eqnarray}
\nonumber 
{\cal G}_{\ell^+ K^+} (j)&=& (1+\Ar)(1+\At)\\ 
\nonumber && \{(1-f_{\Kr}^{++}) [(1-\omega^+) {\cal
  G}_{\Bz\Bz}(j) +\omega^- {\cal G}_{\Bz \Bzb}(j) ] \\ 
\nonumber 
&+& f_{\Kr}^{++}(1-\omega'^{+}){\cal G}_{\Kr}(j) (1+\chi_d \All)~\},\\
\nonumber
{\cal G}_{\ell^- K^-} (j)&=& (1-\Ar)(1-\At) \\
\nonumber && \{(1-f_{\Kr}^{--}) [(1-\omega^-) {\cal
  G}_{\Bzb \Bzb}(j) +\omega^+ {\cal G}_{\Bzb \Bz}(j) ] \\
\nonumber 
&+& f_{\Kr}^{--} (1-\omega'^{-}){\cal G}_{\Kr}(j) (1-\chi_d \All)~\},
\nonumber
\end{eqnarray}}
where the reconstruction asymmetries have separate values
for the $e$ and $\mu$ samples.  We allow for different mistag probabilities
for \Kt ($\omega^\pm$) and \Kr ($\omega'^\pm$).
The parameters $f_{\Kr}^{\pm \pm}(p_k)$ describe the
fractions of \Kr tags in each sample as a function of  
the kaon momentum.  

\begin{center} 
\begin{figure}[htbp]
\includegraphics[width=8cm]{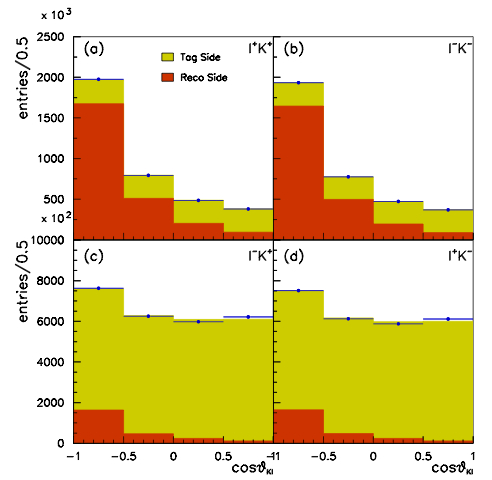}  
\caption{(color online). Distributions of $\cos \theta_{\ell K}$ for the continuum-subtracted 
data (points with error bars) 
and fitted 
contributions from \Brec\ (dark) and \Btag\ (light), for:
(a) $\ell^+ K^+$ events;
(b) $\ell^- K^-$ events; 
(c) $\ell^- K^+$ events; 
(d) $\ell^+ K^-$ events.} 
\label{f:costhe}
\end{figure}
\end{center}

A total of 168 parameters are determined in the fit.
By analyzing 
simulated events as data, we observe that the fit 
reproduces the generated values of $1-|q/p|$ (zero) and of the other most 
significant parameters (\Ar, \At, \dmd, and \tBz). 
We then produce samples of simulated events with 
$\dCP = \pm0.005, \pm0.010, \pm0.025$ and \Ar\ or \At in the range of
$\pm 10\%$, by removing events.
A total of 67 different simulated event samples are used to check for biases.
In each case, the input values are
correctly determined, and an unbiased value of $|q/p|$ is always 
obtained.
\begin{table}[h]
\caption{Principal sources of systematic uncertainties.}
\label{t:syserr}
\begin{center}
\begin{tabular}{l|c}
\hline
Source & $\sigma(\dCP)$ \\
\hline
Peaking Sample Composition&  $^{+1.50}_{-1.17}\times 10^{-3}$\\ 
Combinatorial Sample Composition & $\pm0.39\times 10^{-3}$\\ 
\DT\ Resolution Model & $\pm0.60\times 10^{-3}$\\ 
\Kr\ Fraction & $\pm0.11\times 10^{-3}$\\ 
\Kr\ \DT\ Distribution & $\pm 0.65\times 10^{-3}$\\ 
Fit Bias & $^{+0.58}_{-0.46}\times 10^{-3}$\\ 
\CP eigenstate Description & $\pm 0$\\ 
Physical Parameters & $^{+0}_{-0.28}\times 10^{-3}$\\ 
\hline
\\ [-9.5pt]
Total &  $^{+1.88}_{-1.61}\times 10^{-3}$\\ 
\hline 
\end{tabular}
\end{center}
\end{table}
\begin{center} 
\begin{figure}[htbp]
\includegraphics[width=8cm]{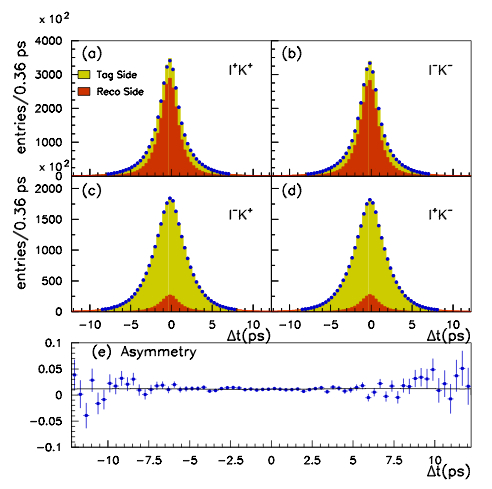}  
\caption{(color online). Distribution of \DT\ for the continuum-subtracted data 
(points with error bars) and fitted 
contributions from \Kr\ (dark) and \Kt\ (light), for:
(a) $\ell^+ K^+$ events;
(b) $\ell^- K^-$ events; 
(c) $\ell^- K^+$ events; 
(d) $\ell^+ K^-$ events; 
(e) raw asymmetry between 
$\ell^+ K^+$ and $\ell^- K^-$ events.}
\label{f:Dz}
\end{figure}
\end{center}
The fit to the data yields 
$\dCP=(0.29\pm0.84^{+1.88}_{-1.61})\times 10^{-3}$, where the first
uncertainty is statistical and the second systematic. 
The values of the detector charge asymmetries are ${\cal A}_{r,e} =( 3.0\pm
0.4) \times 10^{-3}$,  ${\cal A}_{r,\mu}  = (3.1 \pm
0.5)\times 10^{-3}$,  and
$\At = (13.7\pm0.3) \times 10^{-3}$. The frequency of the oscillation
$\dmd=508.5\pm0.9$ \ns$^{-1}$ is consistent
with the world average, while $\tau_{\Bz} = 1.553\pm 0.002$ \ps\  is somewhat
larger than the world average, which we account for in the evaluation of the systematic uncertainties.
Figures \ref{f:Dz} and \ref{f:costhe} show the fit projections 
for \DT\ and $\cos \theta_{\ell K}$. 


The systematic uncertainty is computed 
as the sum in quadrature of several contributions,
described below and summarized in Table~\ref{t:syserr}:
 
- {\it Peaking Sample Composition}: 
we vary the sample composition 
by the statistical uncertainty of the \Mnu\ fit,
the fraction of \Bz\ to \Bp\ in the \dstrstr peaking sample
in the range $50\pm25 \%$ to account for possible violation of isospin symmetry,
the fraction of the peaking contributions (taken from the simulation)
by $\pm20\%$, and the fraction of \CP eigenstates by $\pm 50\% $. 

-{\it \BB\ combinatorial sample composition}: 
we vary the fraction of \Bp\ events in the \BB\ combinatorial sample
by $\pm 4.5\%$, which corresponds to the uncertainty in the
inclusive branching fraction for $\Bz \rightarrow \dsm X$.

-{\it \DT\ resolution model}:   
we quote the difference between the result
when all resolution parameters
   are determined in the fit and those obtained when those that
exhibit a weak correlation with $|q/p|$ are fixed.

-{\it \Kr\ fraction}: 
we vary the ratio of $\Bp \rightarrow \Kr X$ to
$\Bz \rightarrow \Kr X$ by $\pm 6.8\%$, which corresponds to
the uncertainty of the fraction 
$\frac{BR(D^{*0}\rightarrow K^-X)}{BR(D^{*+}\rightarrow K^-X)}$.

-{\it \Kr\ \DT\ distribution}:  
we use half the difference between the results obtained using the two 
different strategies to describe the \Kr\ \DT\ distribution.

-{\it Fit bias}: 
parametrized simulations are used to check the estimate
of the result and its statistical uncertainty. 
We add the statistical uncertainty on the validation test using
the detailed simulation 
and the difference between the nominal result and the central result
determined from the ensemble of parametrized simulations.

-{\it \CP eigenstates description} : 
we vary the $S$ and $C$ parameters describing the \CP eigenstates by their
statistical uncertainties as obtained from simulation.

-{\it Physical parameters}: 
we repeat the fit setting the value of $\Delta\Gamma$ to 0.02 ps$^{-1}$. 
The lifetimes of the $B^0$ and $B^+$ mesons and $\Delta m_d$
are floated in the fit. Alternatively, we check the effect of fixing each 
parameter in turn to the world average.

In summary, we present a new measurement of the parameter governing
\CP violation in \Bz \Bzb oscillations. With a 
partial \Bztodstlnu reconstruction and kaon tagging, we find 
$\dCP = (0.29\pm0.84^{+1.88 }_{-1.61})\times 10^{-3},$ and
$ \ACP = (0.06\pm0.17^{+0.38}_{-0.32})\%.$
These results are consistent with, and more 
precise than, dilepton-based results from B factories~\cite{epem_dilep}. 
No deviation is observed from the SM expectation~\cite{Nierste}.
\begin{acknowledgments}
We are grateful for the excellent luminosity and machine conditions
provided by our \pep2\ colleagues, 
and for the substantial dedicated effort from
the computing organizations that support \babar.
The collaborating institutions wish to thank 
SLAC for its support and kind hospitality. 
This work is supported by
DOE
and NSF (USA),
NSERC (Canada),
IHEP (China),
CEA and
CNRS-IN2P3
(France),
BMBF and DFG
(Germany),
INFN (Italy),
FOM (The Netherlands),
NFR (Norway),
MIST (Russia),
MEC (Spain), and
PPARC (United Kingdom). 
Individuals have received support from the
Marie Curie EIF (European Union) and
the A.~P.~Sloan Foundation.

\end{acknowledgments}
%



\begin{thebibliography}{99}
\bibitem{directCP}B. Aubert {\em et al.} (\babar\ collaboration), 
Phys. Rev. Lett.~{\bf 93},131801(2004).
\bibitem{sin2b}B. Aubert {\em et al.} (\babar\ collaboration), 
Phys. Rev. D~{\bf 79}, 072009 (2009),
 I.Adachi {\em et al.} (Belle collaboration), 
Phys. Rev. Lett.~{\bf 108}, 171802.

\bibitem{Nierste} 
A. Lenz {\em et al.}, Phys. Rev. D~{\bf 86}, 033008 (2012),
J. Charles  {\em et al.}, Phys. Rev. D~{\bf 84}, 033005 (2011),
A. Lenz and U.~Nierste, JHEP {\bf 0706}, 072 (2007).
\bibitem{epem_dilep} B. Aubert {\em et al.} (\babar\ collaboration), 
Phys. Rev. Lett.~{\bf 96}, 251802 (2006),  E. Nakano {\em et al.} (Belle collaboration), 
Phys. Rev. D~{\bf 73}, 112002 (2006).   
\bibitem{D0_mumu} V. M. Abazov {\em et al.} ($D\emptyset$ collaboration), 
Phys. Rev. D~{\bf 84}, 052007 (2011).
\bibitem{D0_B0} V. M. Abazov {\em et al.} ($D\emptyset$ collaboration), 
Phys. Rev. D~{\bf 86}, 072009 (2012).
\bibitem{D0_Bs} V. M. Abazov {\em et al.} ($D\emptyset$ collaboration), 
Phys. Rev. Lett.~{\bf 110}, 011801 (2013).
\bibitem{DKpi} B. Aubert {\em et al.} (\babar\ collaboration),
Phys. Rev. Lett.~{\bf 100}, 051802 (2008).
\bibitem{PDG} J. Beringer {\em et al.}, (Particle Data Group), 
Phys. Rev. D~{\bf 86}, 010001 (2012).
\bibitem{babar_nim} B. Aubert {\em et al.} (\babar\ collaboration),
Nucl. Instr. and Meth. in Phys. Res.~A~{\bf 479}, 1 (2002).
\bibitem{babar_lumi}
J.~P.~Lees {\em et al.} (\babar\ collaboration),
Nucl. Instr. and Meth. in Phys. Res.~A~{\bf 726}, 203 (2013).
\bibitem{evtgen} D.~Lange,
Nucl. Instr. and Meth. in Phys. Res.~A~{\bf 462}, 152 (2001).
\bibitem{wolfram}
G.\ C.\ Fox and S.\ Wolfram, Phys. Rev. Lett. {\bf 41}, 1581 (1978).
\bibitem{BAPX} The \babar\ Physics Book, SLAC Report SLAC-R-504 (1998).
\bibitem{DCS} O. Long, M. Baak, R. N. Cahn and D. Kirkby, 
Phys. Rev. D~{\bf 68}, 034010 (2003).
\end{thebibliography}
\end{document}